\documentclass[a4paper,fleqn]{cas-sc}

\usepackage{amssymb}
\usepackage{amsmath}

\usepackage{graphicx}
\graphicspath{{figs/}}
\usepackage{subcaption}
\usepackage{float}
\usepackage{booktabs}

\usepackage{algorithm}
\usepackage{algpseudocode}

\usepackage{microtype}
\usepackage{xcolor}

\usepackage[authoryear,longnamesfirst]{natbib}
\hypersetup{hypertexnames=false}

\def\tsc#1{\csdef{#1}{\textsc{\lowercase{#1}}\xspace}}
\tsc{WGM}
\tsc{QE}

\begin{document}
\let\WriteBookmarks\relax
\def\floatpagepagefraction{1}
\def\textpagefraction{.001}

\shorttitle{Adaptive Covariance EKF for UWB/PDR Localization}

\shortauthors{Yoo et al.}

\title[mode=title]{KD-EKF: Knowledge-Distilled Adaptive Covariance EKF for Robust UWB/PDR Indoor Localization}

\tnotetext[1]{This research was supported in part by the IITP(Institute of Information \& Coummunications Technology Planning \& Evaluation)-ITRC(Information Technology Research Center) grant funded by the Korea government(Ministry of Science and ICT)(IITP-2026-RS-2021-II211835) and in part by the Korea Institute of Energy Technology Evaluation and Planning (KETEP) and the Ministry of Climate, Energy Environment (MCEE) of the Republic of Korea (No.~RS-2022-KP002860).}

\author{Kyeonghyun Yoo}
\ead{seven1705@korea.ac.kr}

\author{Wooyong Jung}
\ead{jy17347@korea.ac.kr}

\author{Namkyung Yoon}
\ead{nkyoon93@korea.ac.kr}

\author{Sangmin Lee}
\ead{lsm5505@korea.ac.kr}

\author{Sanghong Kim}
\ead{sanghongkim@korea.ac.kr}

\author{Hwangnam Kim}
\cormark[1]
\ead{hnkim@korea.ac.kr}

\affiliation{
  organization={School of Electrical Engineering, Korea University},
  city={Seoul},
  postcode={02841},
  country={Republic of Korea}
}

\cortext[1]{Corresponding author.}

\begin{abstract}
Ultra-wideband (UWB) indoor localization provides centimeter-level accuracy and low latency, but its measurement reliability degrades severely under Non-Line-of-Sight (NLOS) conditions, leading to meter-scale ranging errors and inconsistent uncertainty characteristics. Inertial Measurement Unit (IMU)-based Pedestrian Dead Reckoning (PDR) complements UWB by providing infrastructure-free motion estimation; however, its error accumulates nonlinearly over time due to bias and noise propagation. Fusion methods based on Extended Kalman Filters (EKF) and Particle Filters (PF) can improve average localization accuracy through probabilistic state estimation. However, these approaches typically rely on manually tuned measurement covariances. Such fixed or heuristically tuned parameters are hard to sustain across varying indoor layouts, NLOS ratios, and motion patterns, leading to limited robustness and poor generalization of measurement uncertainty modeling in heterogeneous environments. To address this limitation, this work proposes an adaptive measurement covariance scaling framework in which reliability cues are learned from historical UWB/PDR trajectories. A large teacher model is employed offline to generate temporally consistent next-position predictions from structured UWB/PDR sequences, and this behavior is distilled into a lightweight student model suitable for real-time deployment. The student model continuously regulates EKF measurement covariances based on prediction residuals, enabling environment-aware fusion without manual re-tuning. Experimental results demonstrate that the proposed KD–EKF framework significantly reduces localization error, suppresses error spikes during Line-of-Sight (LOS)/NLOS transitions, and mitigates long-term drift compared to fixed-parameter EKF, thereby improving measurement robustness across diverse indoor environments.
\end{abstract}

\begin{graphicalabstract}
\includegraphics{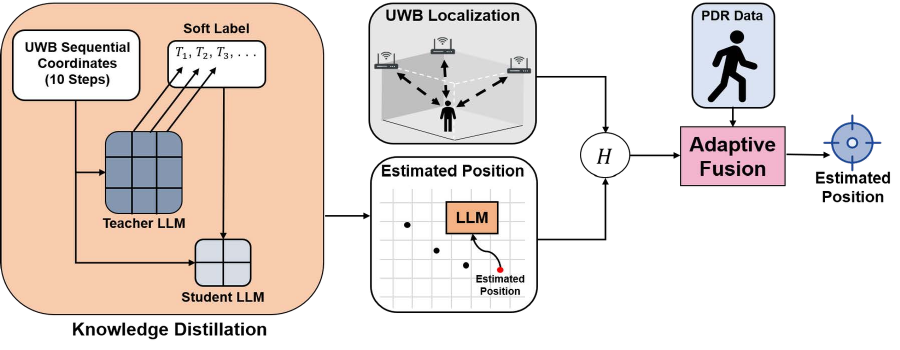}
\end{graphicalabstract}

\begin{highlights}
\item An adaptive EKF-based UWB/PDR fusion framework is proposed to handle time-varying measurement uncertainty.
\item Measurement covariance parameters are recalibrated online to compensate for NLOS-induced ranging degradation and inertial drift.
\item A lightweight reliability estimation model enables real-time uncertainty-aware fusion without manual tuning.
\item Offline knowledge distillation training constructs a compact model while preserving robustness and computational efficiency.
\item Experimental results demonstrate consistent localization accuracy across indoor environments with diverse LOS/NLOS structures.
\end{highlights}


\maketitle

\section{Introduction}
\label{introduction}
The demand for wearable indoor tracking technologies has been rapidly increasing across various domains, including industrial safety monitoring, smart building management, logistics automation, and emergency response~\cite{ROOHI2025111527, BOQUET2024101236, VANHERBRUGGEN2024101260}. These applications require high-precision localization capable of reliably capturing the real-time position of workers or users, making centimeter-level accuracy essential~\cite{983917}. Conventional vision-based or Wi-Fi-based localization methods suffer from environmental dependency and latency issues, whereas Ultra-Wideband (UWB) sensing has emerged as a highly promising technology due to its superior accuracy and responsiveness in indoor environments~\cite{4343996, XU202582, KHAN2025101458}. However, UWB is structurally vulnerable in Non-Line-of-Sight (NLOS) conditions~\cite{10006709, 11120445, HAN2026101062}. Signal reflections, physical obstructions, and multipath propagation can induce meter-scale ranging errors, particularly in corridors or densely structured indoor spaces. A representative complementary approach to mitigate these limitations is Pedestrian Dead Reckoning (PDR), which utilizes wearable Inertial Measurement Unit (IMU) signals. Although PDR provides infrastructure-free displacement estimation based on gait cycles, accumulated sensor bias and noise inevitably lead to significant drift over time~\cite{9162069, Lee2021}.

To address these issues, prior studies have employed fusion techniques such as Extended Kalman Filters (EKF), Particle Filters (PF), or machine-learning-based regression models optimized for specific environments. However, filter-based fusion methods rely on predefined or manually tuned measurement covariance models. When noise statistics or environmental conditions change, these fixed assumptions often require repeated manual adjustment, limiting robustness across heterogeneous indoor environments~\cite{8954658, 9531633, HABASH2024101388}. Machine Learning (ML) models, in contrast, tend to overfit dataset-specific distributions, resulting in poor generalization under heterogeneous environments~\cite{NGUYEN2024101912, 11077686, AZADDEL2023100967}.

Recently, pretrained sequence models, including instruction-tuned LLMs, have shown that they can act as general-purpose pattern extractors when sensor trajectories are represented in a structured textual form. In our setting, the teacher model is not used to interpret linguistic semantics, but to provide a temporally consistent next-position prediction that reflects trajectory-level regularities such as motion continuity, directional changes, and drift-like deviation behavior. Large discrepancies between UWB measurements and the teacher-consistent prediction are regarded as indicators of degraded ranging quality, which in turn drive dynamic covariance scaling within the EKF framework~\cite{10831691, Abdullah2025, 9795287}. This reasoning capability suggests a new direction for addressing the environmental uncertainty and non-stationary sensor patterns that traditional fusion methods struggle to handle. In particular, such capabilities can be exploited to model time-varying measurement reliability and uncertainty, which are difficult to capture using fixed or heuristic covariance models~\cite{10006709, SKRZEK2026101045}.
However, LLMs incur substantial computational overhead and latency, making them impractical for real-time use on wearable or mobile edge devices~\cite{11141409, GONG2026101012, LI2024100612, MA2025100832, 10852525}.

Therefore, robust indoor localization requires explicit modeling of time-varying measurement reliability and uncertainty rather than relying on fixed noise assumptions. This work proposes an adaptive fusion framework that dynamically estimates UWB ranging degradation and PDR drift uncertainty and corrects the EKF measurement covariance in real time.

To efficiently learn complex and non-stationary uncertainty patterns from multimodal sensing data, a large-scale teacher model is employed during offline training, and its reliability inference capability is transferred to a lightweight student network through Knowledge Distillation (KD). The student model estimates measurement reliability indicators at each timestep and injects them into the EKF by adaptively scaling the measurement covariance matrix $\mathbf{R}$ and fusion weights, enabling robust localization without manual tuning across heterogeneous indoor environments.

The main contributions of this paper are as follows:

\begin{enumerate}
    \item This work formulates UWB/PDR fusion not as a fixed-covariance sensor-weighting scheme, but as a dynamic measurement reliability estimation problem, explicitly incorporating NLOS-induced ranging errors and accumulated PDR drift into the estimation structure.

    \item A regression-based inference mechanism is introduced to predict the next position from historical trajectory sequences, where the resulting prediction error is utilized as a quantitative indicator of measurement reliability, integrating time-series position forecasting with reliability estimation.

    \item A real-time covariance adjustment mechanism is designed to directly inject the estimated reliability indicators into the EKF measurement covariance matrix, enabling adaptive fusion without manual parameter retuning under varying environmental conditions.

    \item In the offline stage, a large-scale Teacher model is employed to learn complex non-stationary temporal patterns, and its inference capability is transferred to a lightweight Student model through KD, resulting in a KD-EKF architecture suitable for real-time deployment in embedded environments.
\end{enumerate}

The remainder of this paper is organized as follows.  
Section~\ref{related_work} reviews indoor localization and sensor fusion techniques based on UWB and PDR, and analyzes the limitations of fixed noise models and manually tuned approaches in capturing time-varying measurement uncertainty under changing environmental conditions.  
Section~\ref{system} presents the proposed KD-EKF adaptive sensor fusion framework, which recalibrates the EKF measurement covariance matrix in real time using an offline-trained lightweight reliability estimation model.  
Section~\ref{eval} evaluates the accuracy, robustness, and real-time feasibility of the proposed approach through experiments conducted in indoor environments with different Line-of-Sight (LOS)/NLOS transition characteristics, and quantitatively analyzes the individual contributions of KD, lightweight model deployment, and adaptive covariance correction via ablation studies.  
Finally, Section~\ref{con} concludes the paper and discusses future research directions.

\section{Related Work}
\label{related_work}
Indoor localization remains challenging due to nonlinear sensor behavior, environmental variability, and frequent NLOS propagation. UWB ranging achieves high accuracy under LOS conditions, but signal blockage and reflection in NLOS environments introduce large ranging biases. Prior studies addressed this issue by classifying UWB measurements into LOS/NLOS categories and compensating for excess delays before position estimation, thereby improving robustness under adverse channel conditions. Meanwhile, PDR provides infrastructure-free displacement estimation but suffers from cumulative drift caused by IMU bias, thermal variation, and device-dependent characteristics, which limits its long-term reliability. As a result, standalone UWB or PDR systems struggle to maintain stable localization performance in dynamic indoor environments~\cite{10006709, 11120445, HAN2026101062, 9611295, Lee2021, 4343996, XU202582, KHAN2025101458}.

To overcome these limitations, probabilistic fusion of UWB and IMU sensing has been widely studied. EKF and PF tightly couple inertial motion prediction with UWB ranging updates, achieving higher accuracy than least-squares methods while remaining suitable for embedded implementations. However, Kalman filter-based approaches rely on manually tuned covariance matrices, making them sensitive to environment-dependent noise characteristics. Adaptive filtering methods, such as Mahalanobis-distance-based outlier detection, partially alleviate this issue by adjusting measurement uncertainty during NLOS conditions. Nevertheless, these techniques still depend on heuristic parameter tuning and exhibit limited generalization across diverse environments.
In parallel, learning-based indoor localization has advanced through the use of Wi-Fi fingerprints, IMU sequences, and UWB Channel Impulse Responses (CIR). Neural models based on convolutional architectures and CIR features have shown enhanced capability in mitigating ranging bias under controlled conditions. Transformer-based architectures further enhance sequential modeling capability but incur substantial computational overhead, which restricts real-time deployment on lightweight edge platforms~\cite{8954658, 9162069, 10831691, 9531633, 10526268, NGUYEN2024101912, Abdullah2025}.

Recent studies have explored data-driven approaches to more accurately reflect measurement quality variations arising from complex indoor sensing conditions, such as NLOS transitions, multipath distortion, and cumulative inertial sensor drift. These models aim to predict the reliability of UWB ranging and IMU-based motion estimation by analyzing sequential sensor observations, thereby overcoming the limitations of conventional probabilistic fusion methods that rely on fixed noise assumptions. However, high-capacity sensor interpretation models typically involve substantial computational overhead and memory requirements, which hinder their direct deployment in real-time wearable and embedded localization systems~\cite{9795287, 11141409, Rizwan2025, 11141409, GONG2026101012, LI2024100612, MA2025100832, 10852525}.

To address this limitation, this work adopts a two-stage learning and deployment strategy. During the offline training phase, a large-scale prediction model is designed to perform measurement quality estimation tasks based on multimodal sensor sequences, producing reliability indicators for UWB and PDR observations. These outputs are then used as supervisory signals to transfer the predictive capability to a lightweight student network through KD. During online operation, only the computationally efficient student model is employed to estimate measurement reliability in real time, and the inferred reliability information is injected into the EKF by dynamically adjusting the measurement covariance matrix $\mathbf{R}$ and fusion weighting. This framework enables adaptive sensor fusion under time-varying environmental conditions while maintaining real-time feasibility on resource-constrained platforms.

\section{System Design}
\label{system}

\begin{figure}
\centering
\includegraphics[height=7.5cm, keepaspectratio]{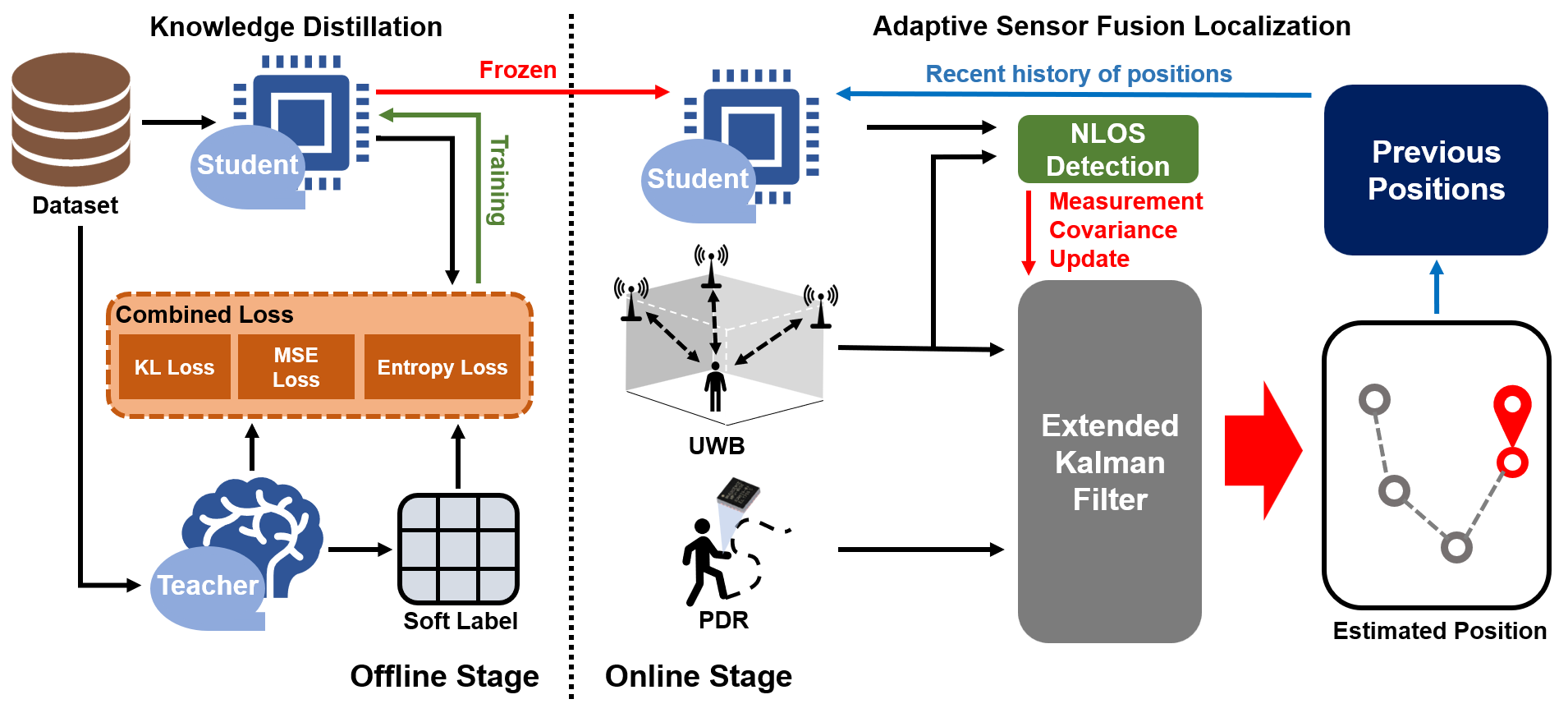}
\caption{Two-stage system overview of the proposed KD–EKF hybrid localization framework.}
\label{fig1}
\end{figure}

The proposed system reformulates indoor UWB/PDR fusion as a dynamic measurement reliability and uncertainty estimation problem rather than a conventional state estimation task. Instead of directly combining sensor outputs through fixed or heuristically tuned filtering mechanisms, the framework explicitly infers the evolving reliability of UWB measurements under NLOS-induced degradation and inertial drift effects, and incorporates this reliability into probabilistic state correction via adaptive covariance regulation. This real-time covariance adaptation mechanism constitutes the core operational link between learned reliability inference and probabilistic sensor fusion, enabling continuous correction of EKF measurement uncertainty without manual parameter tuning.

As illustrated in Fig.~\ref{fig1}, the framework consists of an offline stage and an online EKF-based fusion stage. In the offline stage, a temporally consistent next-position estimation model is constructed from historical UWB and PDR position sequences, providing trajectory-aware predictions that serve as a reliability reference for subsequent uncertainty modeling.

In the online stage, the discrepancy between the model-based position prediction and the incoming UWB measurement is exploited to assess measurement quality and to dynamically recalibrate the EKF measurement covariance in real time. This section presents the UWB/PDR positioning modules and details the proposed KD–EKF-based adaptive fusion process.
\subsection{UWB Positioning using DS-TWR Trilateration}
The UWB positioning module adopts the Double-Sided Two-Way Ranging (DS-TWR) method between fixed anchors and a mobile tag~\cite{10831691}. When the tag transmits an initial poll signal, an anchor receives it and then sends back a response signal. The tag measures the round-trip Time-of-Flight (TOF) and compensates for clock offset using the anchor’s response latency, thereby obtaining the distance $r_i$ to the $i$-th anchor. Given distance measurements from $n$ anchors, the 2D position $(x,y)$ of the tag is obtained via the following trilateration equations:
\[
(x - x_i)^2 + (y - y_i)^2 = r_i^2, \qquad i = 1, \dots, n,
\]
where $(x_i,y_i)$ denotes the position of the $i$-th anchor. At least three anchors in 2D and four in 3D are required, and the nonlinear equations are solved using Gauss–Newton linearization. 

While DS-TWR provides centimeter-level accuracy under LOS conditions, NLOS environments may introduce several meters of ranging error due to signal blockage and reflections, which increase propagation delay. Within the EKF, UWB measurement errors are modeled using a baseline measurement covariance $R_0$, which is later scaled dynamically during the online stage according to prediction errors.

\subsection{PDR using Chest-Mounted IMU}
The PDR module tracks pedestrian motion using a chest-mounted IMU consisting of a 3-axis accelerometer and a gyroscope. The overall process consists of the following three steps.
\begin{enumerate}
\item \textbf{Step detection}: Periodic peaks in the vertical acceleration signal are detected using a threshold-based method to segment individual gait cycles.
\item \textbf{Step length estimation}: The step length $s_k$ is estimated using the Weinberg method, which relates acceleration amplitude and gait period:
\begin{equation}
s_k = K \cdot \sqrt[4]{a_{\mathrm{max},k} - a_{\mathrm{min},k}},
\end{equation}
where $K$ is a calibration constant determined by the user’s gait characteristics, and $a_{\mathrm{max},k}$ and $a_{\mathrm{min},k}$ denote the maximum and minimum vertical accelerations in the $k$-th gait cycle~\cite{weinberg2002using}.

\item \textbf{Heading estimation}: The walking direction $\theta_k$ is obtained by integrating the gyroscope signal. Since magnetometers are highly susceptible to interference in indoor environments, no separate magnetic calibration is applied, and heading drift is automatically corrected through UWB updates within the EKF.
\end{enumerate}

The 2D position of the pedestrian is then updated as
\begin{equation}
\mathbf{p}_{k+1} 
= 
\mathbf{p}_{k} +
s_{k}\,\bigl(\cos\theta_{k}, \sin\theta_{k}\bigr),
\end{equation}
where $\mathbf{p}_{k}$ is the current position, $s_k$ is the estimated step length, and $\theta_k$ is the walking direction obtained from gyroscope integration. Although PDR errors accumulate over time due to IMU drift, such drift is suppressed in LOS segments via UWB-based EKF updates.

\subsection{Offline Stage: LLM Teacher-Student Knowledge Distillation}
\label{offline_kd}
In the offline stage, knowledge is distilled from a pretrained large-scale LLM-based teacher into a lightweight student model using the teacher’s inference outputs. The teacher remains fixed during training, while the student is optimized to mimic its outputs. The resulting student model is deployed in the online stage to predict positions and to provide a reliability signal used for adaptive scaling of the EKF measurement covariance \textit{R}.

\subsubsection{Teacher LLM for Position Regression}
\label{teacher_reg}
The teacher model is built upon a pretrained LLaMA-family large language model and performs next-position regression using a structured prompt constructed from recent UWB/PDR time-series data. The prompt represents motion dynamics by providing the most recent $L$ two-dimensional position coordinates in chronological order, along with lightweight trend summaries of the $x$-axis and $y$-axis trajectories categorized as increasing, decreasing, or stationary. These low-dimensional summaries, directly derived from observed coordinates, encourage the model to capture global motion directionality rather than overreacting to local fluctuations.

By jointly leveraging coordinate history and trend information, latent motion patterns—such as walking speed changes, turning behavior, and trajectory curvature—are implicitly encoded, enabling trajectory-level reasoning that favors continuity and spatial flow over naive coordinate extrapolation. The prompt ends with an explicit request for the next position, encouraging the model to jointly account for local temporal dependencies and global spatial context while maintaining a concise, noise-free formulation.

Both the teacher and student models use an identical prompt format, ensuring a shared temporal representation through which the student learns to mimic the teacher’s trajectory interpretation without relying on prompt-specific heuristics. The teacher outputs the predicted next position $(x_{k+1}, y_{k+1})$ in textual form, which is parsed into a real-valued vector $\hat{\mathbf{p}}_{\mathrm{t}}$ using a regular-expression-based parser and used as a soft target for training the student model.

\subsubsection{Knowledge Distillation}
During student training, the teacher LLM performs prompt-based inference for each input sequence and generates a real-time prediction $\hat{\mathbf{p}}_{\mathrm{t}}$ for the next position. Through KD, the student model adopts the teacher’s coordinate predictions as soft labels and learns to approximate the teacher’s inference behavior. The KD process is carried out without updating the teacher’s parameters and allows the student to internalize the teacher’s sequence interpretation and spatial reasoning patterns in a structured manner. As a result, the student model acquires predictive capability that reflects the overall structure and dynamics of the trajectory, going beyond simple coordinate regression.
\subsubsection{Lightweight Student Model}
The student model is built upon a lightweight LLaMA-family language model with approximately one billion parameters and employs the same tokenizer as the teacher model. This design preserves consistency in prompt structure and token-level representations, enabling stable transfer of the distributional knowledge embedded in the teacher’s output probability distribution, referred to as soft labels. The student model receives the same prompt as the teacher and predicts the next position; during training, it learns to imitate the teacher-provided soft labels, thereby effectively capturing the teacher’s inference behavior.

During training, the student is supervised by multiple signals produced by the teacher, including the soft token distribution and the teacher’s predicted coordinates. The student is optimized by minimizing the following composite loss:
\begin{equation}
\mathcal{L}_{\mathrm{total}}
=
\alpha \, \mathcal{L}_{\mathrm{MSE}}
+
\beta \, \mathcal{L}_{\mathrm{KD}}
+
\gamma \, \mathcal{L}_{\mathrm{text}},
\end{equation}
where the weighting factors are empirically set to $\alpha = 0.5$, $\beta = 0.3$, and $\gamma = 0.2$ to balance geometric accuracy, distributional imitation, and output consistency.

The student model is trained using a composite loss consisting of three complementary terms, each designed to transfer a different aspect of the teacher’s knowledge.

\begin{itemize}
    \item \textbf{Logit-Based Distillation Loss ($\mathcal{L}_{\mathrm{KD}}$):}
    To transfer the teacher’s distributional knowledge, the student minimizes the Kullback-Leibler (KL) divergence between the teacher and student output distributions at the final token.  Let $z_{\mathrm{t}}$ and $z_{\mathrm{s}}$ denote the teacher and student logits, and $p_{\mathrm{t}}$, $p_{\mathrm{s}}$ their corresponding softmax distributions. The loss is defined as
    \begin{equation}
    \mathcal{L}_{\mathrm{KD}}
    =
    D_{\mathrm{KL}}
    \!\left(
    p_{\mathrm{t}}
    \;\Vert\;
    p_{\mathrm{s}}
    \right).
    \end{equation}
    For computational efficiency, the KL divergence is computed only over the top-$K$ tokens of the teacher distribution.

    \item \textbf{Coordinate Regression Loss ($\mathcal{L}_{\mathrm{MSE}}$):}
    In addition to token-level supervision, the student is guided in the coordinate space through a lightweight regression head.  
    The final hidden state $\mathbf{h}_{\mathrm{s}}$ is mapped to a two-dimensional position estimate
    \begin{equation}
    \hat{\mathbf{p}}_{\mathrm{s}} = f_{\mathrm{coord}}(\mathbf{h}_{\mathrm{s}}),
    \end{equation}
    and the regression loss is defined as
    \begin{equation}
    \mathcal{L}_{\mathrm{MSE}}
    =
    \left\|
    \hat{\mathbf{p}}_{\mathrm{s}}
    -
    \hat{\mathbf{p}}_{\mathrm{t}}
    \right\|_2^2,
    \end{equation}
    where $\hat{\mathbf{p}}_{\mathrm{t}}$ denotes the teacher’s predicted coordinates.  
    This term encourages the student to imitate the teacher’s high-level trajectory patterns.

    \item \textbf{Text-Based Consistency Loss ($\mathcal{L}_{\mathrm{text}}$):}
To enforce consistency between linguistic generation and numerical regression, an optional text-based consistency loss is introduced.  
Let $\hat{\mathbf{p}}^{\text{text}}_s \in \mathbb{R}^2$ denote the coordinates parsed from the student model’s generated textual output, and let $\hat{\mathbf{p}}_t \in \mathbb{R}^2$ represent the teacher’s predicted coordinates.  
The loss is defined as
\begin{equation}
\mathcal{L}_{\mathrm{text}}
=
\left\|
\hat{\mathbf{p}}^{\text{text}}_s
-
\hat{\mathbf{p}}_t
\right\|_2^2 ,
\end{equation}
which encourages alignment between textual reasoning and regression outputs, thereby improving robustness in prompt-based inference scenarios.
\end{itemize}

\subsection{Online Stage: Error-Based Adaptive Sensor Fusion}
\label{online_fusion}

In the online stage, the proposed framework interprets the discrepancy between learned trajectory priors and real-time UWB observations as an indicator of measurement reliability, and formulates sensor fusion as a dynamic uncertainty estimation process embedded within the EKF. The EKF state is defined as
\begin{equation}
\mathbf{x}_k = [p_x, p_y, v_x, v_y]^\top,
\end{equation}
where $(p_x, p_y)$ denotes position and $(v_x, v_y)$ denotes velocity. The state is propagated using the velocity obtained from PDR and updated with UWB trilateration measurements.

At each time step $k$, the student LLM predicts the next position $\hat{\mathbf{x}}^{\mathrm{LLM}}_k$ from the recent UWB/PDR sequence. The discrepancy between the UWB measurement $\mathbf{z}_k$ and the student prediction is defined as
\begin{equation}
\Delta_k = \left\| \mathbf{z}_k - \hat{\mathbf{x}}^{\mathrm{LLM}}_k \right\|.
\end{equation}
A small $\Delta_k$ indicates agreement between the UWB measurement and the LLM-based prediction, whereas a large $\Delta_k$ suggests potential NLOS conditions or reduced UWB reliability.

A sliding window $\Delta_{\mathrm{win}}$ of recent errors is used to compute the median and median absolute deviation (MAD), based on which the following dynamic thresholds are defined:
\begin{align}
\theta_1 &= \mathrm{median} + \alpha \times \mathrm{MAD},\\
\theta_2 &= \mathrm{median} + \beta \times \mathrm{MAD}.
\end{align}
The scaling factors are denoted as $\alpha$ and $\beta$, corresponding to progressively larger deviations with respect to the nominal error distribution, where the specific values $\alpha = 1.5$ and $\beta = 3.0$ are selected as empirically grounded yet conservative settings representing moderate and significant departures from typical measurement noise, respectively. This parameterization enables a soft reliability transition without relying on hard binary NLOS classification.

Based on the error level, the scaling factor $H_k$ is selected as follows:
\begin{itemize}
    \item If $\Delta_k \le \theta_1$: $H_{\mathrm{low}}$ (high UWB reliability)
    \item If $\theta_1 < \Delta_k \le \theta_2$: $H_{\mathrm{mid}}$ (moderate degradation)
    \item If $\Delta_k > \theta_2$ for $n$ consecutive steps: $H_{\mathrm{high}}$ (persistent NLOS)
\end{itemize}

The UWB measurement covariance \textit{R} is then adjusted as
\begin{equation}
R_k = H_k \cdot R_0,
\end{equation}
where $R_0$ denotes the baseline LOS covariance. When the error is small, the EKF places greater trust in UWB measurements to suppress PDR drift; when the error is large, it reduces the influence of potentially biased UWB values and relies more on PDR prediction.

The online adaptive scaling procedure is summarized in Algorithm~\ref{llm_ekf}. Through this real-time covariance adaptation mechanism, the proposed framework continuously injects learned reliability indicators into the EKF measurement covariance $\mathbf{R}$ and fusion weighting. This dynamic correction process directly governs fusion behavior under heterogeneous sensing conditions, enabling robust localization without reliance on static noise assumptions or manual retuning. In particular, the use of MAD-based dynamic thresholds computed over a sliding window of the most recent 10 error samples allows the system to remain robust against sudden outliers while automatically adjusting the relative trust between UWB and PDR according to changes in indoor conditions.

\section{Evaluation}
\label{eval}
This section evaluates the effectiveness of the proposed LLM-based KD adaptive fusion framework for UWB/PDR-based indoor localization. The evaluation focuses on localization accuracy and robustness under varying sensor quality and NLOS conditions. Quantitative comparisons are conducted against EKF-based UWB/PDR fusion frameworks that rely on fixed covariance assumptions. The experimental environment configuration, evaluation metrics, and quantitative results are presented in detail in the following subsections.

\begin{algorithm}
\caption{LLM-Based Adaptive Measurement Covariance Scaling for EKF}
\label{llm_ekf}
\begin{algorithmic}[1]

\Require UWB measurement $\mathbf{z}_k$, LLM prediction $\hat{\mathbf{x}}^{\mathrm{LLM}}_k$, baseline covariance $R_0$
\Ensure Updated covariance $R_k$, NLOS flag

\Statex \textbf{Parameters:} error window $\Delta_{\mathrm{win}}$; thresholds $(\alpha,\beta)$;
scaling factors $(H_{\mathrm{low}}, H_{\mathrm{mid}}, H_{\mathrm{high}})$; persistence length $n$

\State $NLOS \gets \textbf{False}$; $R_k \gets R_0$
\While{system is running}
    \State $\Delta_k \gets \lVert \mathbf{z}_k - \hat{\mathbf{x}}^{\mathrm{LLM}}_k \rVert_2$
    \State Append $\Delta_k$ to $\Delta_{\mathrm{win}}$
    \State Compute median and MAD of $\Delta_{\mathrm{win}}$
    \State $\theta_1 \gets \mathrm{median}(\Delta_{\mathrm{win}}) + \alpha \cdot \mathrm{MAD}(\Delta_{\mathrm{win}})$
    \State $\theta_2 \gets \mathrm{median}(\Delta_{\mathrm{win}}) + \beta \cdot \mathrm{MAD}(\Delta_{\mathrm{win}})$
    \If{$\Delta_k > \theta_2$ for $n$ consecutive frames}
        \State $NLOS \gets \textbf{True}$
        \State $H_k \gets H_{\mathrm{high}}$
    \ElsIf{$\Delta_k > \theta_1$}
        \State $H_k \gets H_{\mathrm{mid}}$
    \Else
        \State $H_k \gets H_{\mathrm{low}}$
    \EndIf
    \State $R_k \gets H_k \cdot R_0$
    \State Update EKF using $R_k$
\EndWhile

\end{algorithmic}
\end{algorithm}
\subsection{Experimental Setup}
\label{exp_setup}
A distributed experimental architecture was adopted to evaluate the proposed KD-EKF framework under realistic sensing and computation constraints. UWB-based ranging and IMU-based PDR preprocessing were executed on an embedded platform, while high-level position prediction and sensor fusion were performed on a server. Step detection, step length estimation, and heading computation were implemented on a Raspberry Pi~4B (8GB), and the preprocessed data were transmitted to a server equipped with an AMD Ryzen Threadripper PRO~5955WX CPU and two NVIDIA RTX~4080 GPUs.

UWB measurements were collected using a Qorvo DWM1000 module, and IMU data were obtained from an MPU6050 sensor at a sampling rate of 100\,Hz. Ground-truth (GT) trajectories were established by manually measuring reference points and distances within the indoor environment. Sensor fusion was realized through an EKF-based framework with the state vector defined as $\mathbf{x}_k = [p_x, p_y, v_x, v_y]^\top$.

For position prediction, the teacher and student models were implemented using LLaMA-3.1-8B-Instruct and LLaMA-3.2-1B-Instruct, respectively, sharing an identical tokenizer. Prompt-based position regression was conducted with an input sequence length of $L=10$, where each prompt consisted of the ten most recent two-dimensional positions and the model predicted the position at the next time step. The KD dataset was constructed from eleven independent trajectory cases, yielding 7{,}000 samples of eleven consecutive positions, among which 6{,}000 samples were used for training and 1{,}000 samples were reserved for evaluation. During online operation, only the lightweight student model was deployed, performing real-time inference at each PDR time step.

\begin{figure}
\centering
\includegraphics[width=1.0\linewidth]{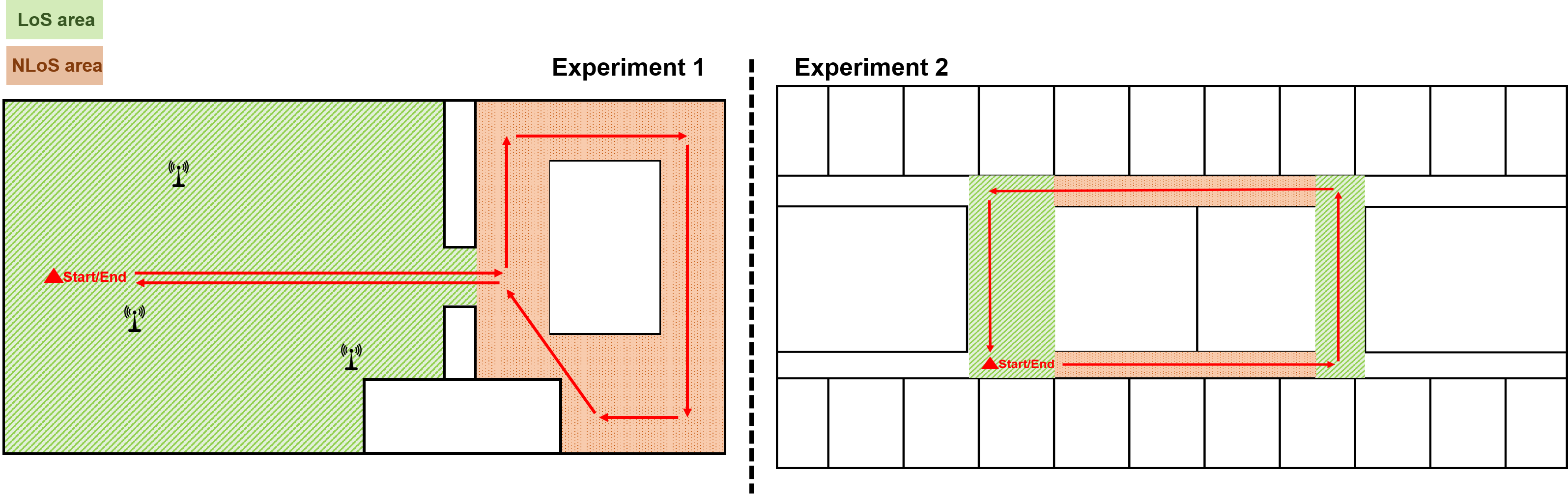}
\caption{Floor plans and experimental trajectories used in the evaluation.
Green areas denote LOS areas covered by UWB anchors, while orange areas
indicate NLOS areas without anchor coverage.
Experiment~1 represents a short-range environment with separated LOS/NLOS
segments, whereas Experiment~2 corresponds to a closed rectangular corridor
with repeated LOS/NLOS transitions.}
\label{exp_layout}
\end{figure}

All reported performance metrics were computed after applying a unified post-processing step to suppress high-frequency estimation oscillations. Fig.~\ref{exp_layout} illustrates the indoor environments and the predefined walking trajectories used in the experiments. In the figure, LOS areas are defined as regions where UWB anchors are deployed
and signal visibility is ensured, and are indicated as green hatched areas,
while NLOS areas correspond to segments where anchors are not installed or
where signals are blocked or reflected by walls and structural obstacles, and are shown as orange dashed areas. The two environments were intentionally designed to exhibit different LOS/NLOS transition patterns in order to evaluate the environment generalization capability of the proposed KD-EKF framework.

\begin{figure}
\centering
\begin{subfigure}{0.55\linewidth}
    \centering
    \includegraphics[width=\linewidth,trim=0 0 2 0,clip]{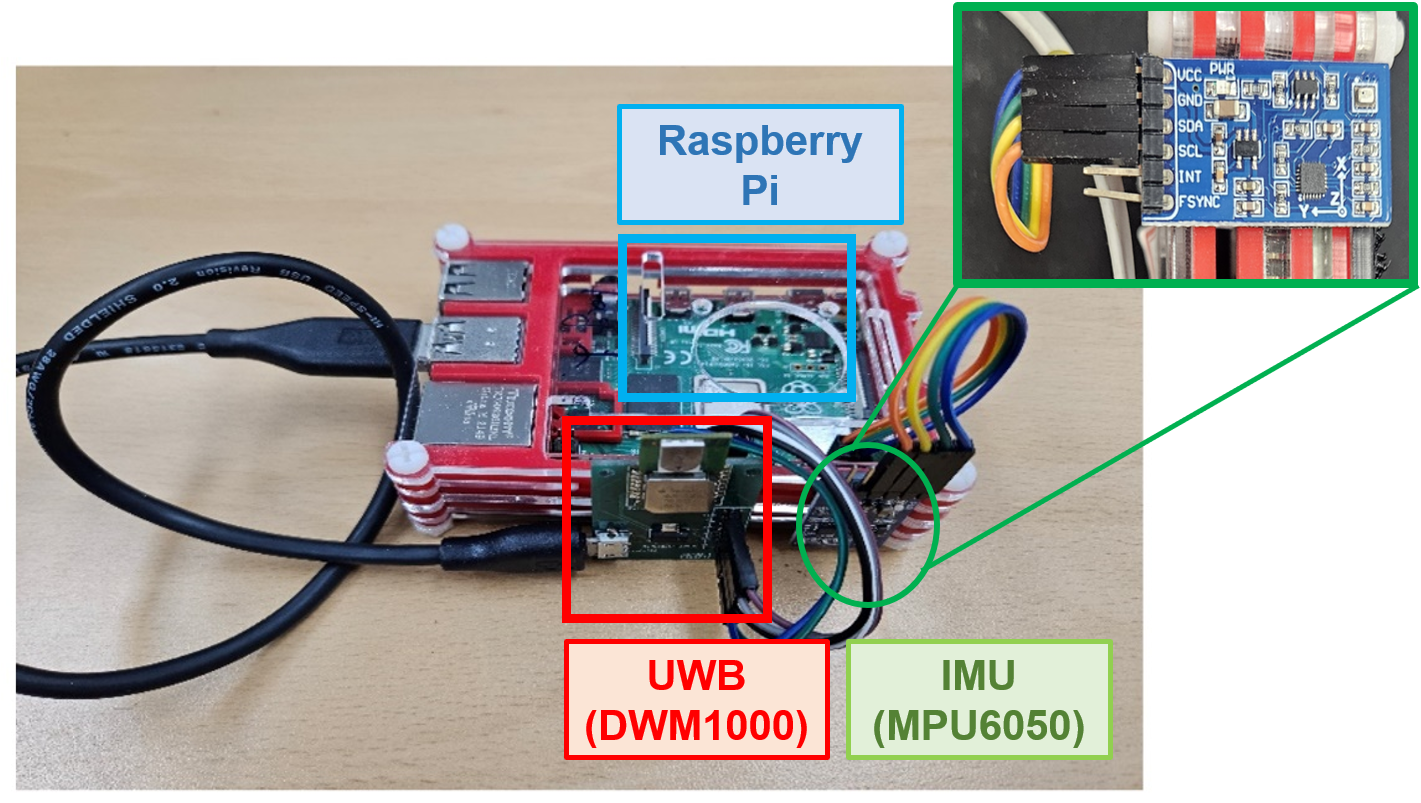}
    \caption{Mobile UWB/PDR tag device}
    \label{tag}
\end{subfigure}
\hfill
\begin{subfigure}{0.43\linewidth}
    \centering
    \includegraphics[width=\linewidth,trim=2 0 0 0,clip]{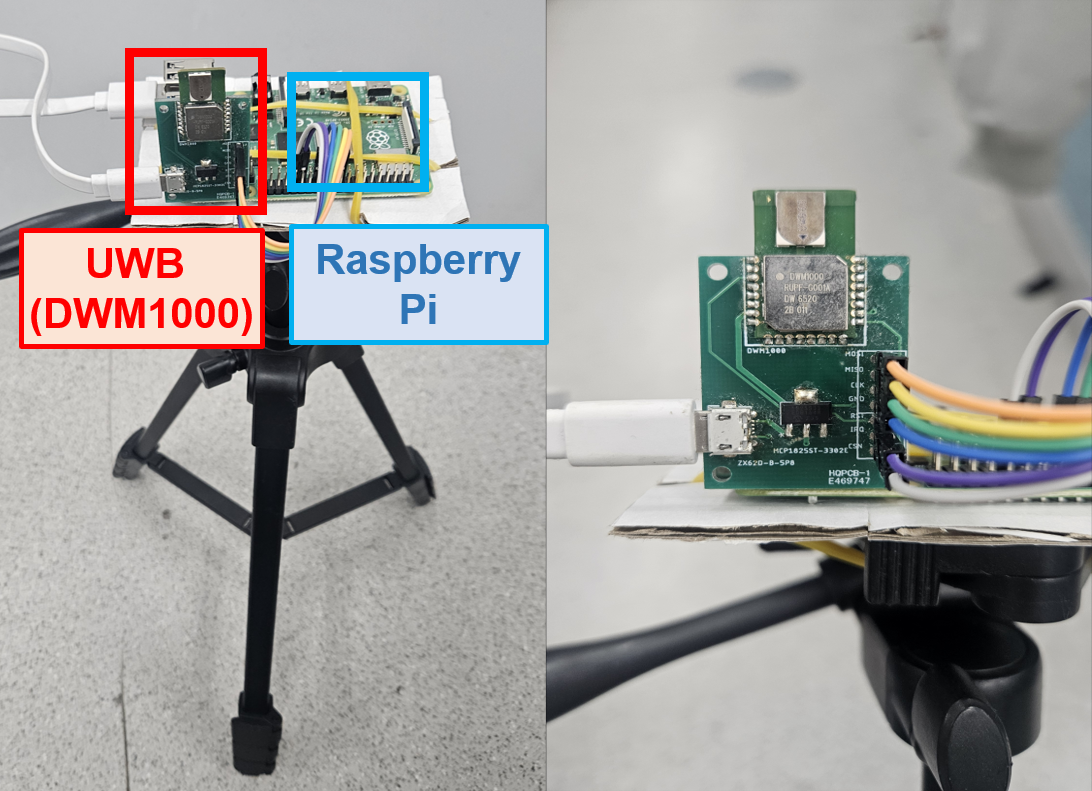}
    \caption{Fixed UWB anchor device}
    \label{anchor}
\end{subfigure}
\caption{Experimental hardware configuration.
(a) Mobile tag device integrating a Raspberry Pi, a DWM1000 UWB module,
and an IMU sensor for PDR preprocessing.
(b) Fixed UWB anchor device based on the DWM1000 module deployed in LOS areas.}
\label{hardware}
\end{figure}

Fig.~\ref{hardware} shows the experimental hardware used for data collection. The mobile tag integrates UWB and IMU sensors to enable synchronized ranging and PDR measurements, while multiple fixed anchors are deployed at predefined
locations to provide LOS-based UWB ranging references.

\subsection{Prompt-Based Position Regression Performance and Knowledge Distillation Effect}
\label{llm_regression}

This experiment evaluates the effectiveness of the prompt-based position regression scheme described in Section~\ref{teacher_reg} and investigates how well the teacher’s trajectory interpretation capability is transferred to the lightweight student model via the KD framework in Section~\ref{offline_kd}. The teacher, student, and base models were evaluated using identical prompts and unseen GT trajectory segments, with the same input sequence length ($L=10$) and inference rate across all conditions.

Quantitative performance was assessed using the RMSE of Absolute Trajectory Error (ATE), defined as
\begin{equation}
\mathrm{ATE}_{\mathrm{RMSE}}
=
\sqrt{
\frac{1}{N}
\sum_{k=1}^{N}
\left\|
\mathbf{p}_k^{\mathrm{est}}
-
\mathbf{p}_k^{\mathrm{gt}}
\right\|_2^2
},
\end{equation}
where $\mathbf{p}_k^{\mathrm{est}}$ and $\mathbf{p}_k^{\mathrm{gt}}$ denote the estimated and ground-truth positions at time step $k$, respectively.

As summarized in Table~I, the teacher model achieved an ATE$_{\mathrm{RMSE}}$ of 0.3067\,m. The student model trained with KD reduced the error to 1.2311\,m, whereas the base student model without KD exhibited a higher error of 1.8713\,m. Differences in LLaMA model versions may affect performance due to variations in training data and alignment, but the student model was selected in the 1B-parameter regime with the same tokenizer as the teacher to ensure stable distillation. These results indicate that the observed performance gain is primarily attributable to KD rather than model version differences.

\begin{figure}
\centering
\includegraphics[width=0.85\linewidth]{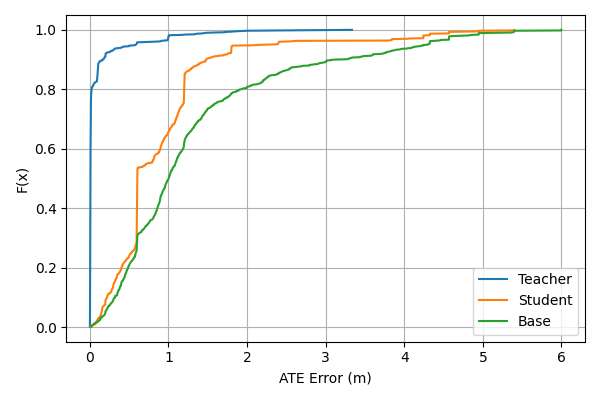}
\caption{CDF of Absolute Trajectory Error (ATE)}
\label{ate_cdf}
\end{figure}

Fig.~\ref{ate_cdf} further illustrates the error distributions. The student model shows a clear leftward shift of the CDF compared to the base student model, indicating a more stable and consistent error profile. Although a performance gap remains between the teacher and student models due to differences in architectural capacity, KD markedly suppresses large-error events in the student model, indicating that KD mainly enhances regression stability rather than average accuracy alone.

These results demonstrate that the proposed prompt-based regression structure enables LLMs to interpret temporal position sequences as a continuous regression task, and that KD effectively transfers this capability to a lightweight model suitable for real-time EKF-based sensor fusion.

\subsection{Teacher-Student Inference Latency and Throughput Comparison}
\label{latency_throughput}
This experiment aims to verify the necessity of KD in real-time indoor localization systems by quantitatively comparing the inference latency and throughput of a large-scale teacher model and a lightweight student model. The comparison was conducted under identical input prompts and execution environments, where the end-to-end inference latency was measured as the elapsed time from prompt input to the final position output, and throughput was defined as the number of inferences processed per second (Hz). The evaluation included a teacher model (LLaMA-3.1-8B), a student model trained through KD (LLaMA-3.2-1B), and a base student model with the same architecture but trained without KD. All models were evaluated under the same single-GPU setting.

As shown in Table~\ref{latency}, the teacher model exhibited an average inference latency of 283.1\,ms and a throughput of 3.5\,Hz, indicating that it struggles to meet the tens-of-Hz throughput required for real-time EKF-based sensor fusion. In contrast, the student model achieved an average latency of 129.2\,ms and a throughput of 7.7\,Hz, corresponding to approximately 2.2$\times$ lower latency and 2.2$\times$ higher throughput compared to the teacher model. Meanwhile, the base student model trained without KD showed nearly identical inference performance, with a latency of 131.5\,ms and a throughput of 7.6\,Hz. This observation indicates that KD does not alter the student model’s architecture or inference pathway and therefore does not impose a noticeable overhead on inference latency or throughput. These results demonstrate that KD effectively improves regression accuracy without sacrificing inference efficiency, thereby enabling a favorable trade-off between computational cost and performance for real-time EKF-based indoor localization systems.
\begin{table}
\centering
\caption{Comparison of Position Regression Performance (ATE$_{\mathrm{RMSE}}$)}
\begin{tabular}{c c}
\hline
Model & ATE$_{\mathrm{RMSE}}$ (m) \\
\hline
Teacher model (LLaMA-3.1-8B) & 0.3067 \\
Student model (LLaMA-3.2-1B, KD) & 1.2311 \\
Base student model(LLaMA-3.2-1B) & 1.8713 \\
\hline
\end{tabular}
\label{llm_ate}
\end{table}
\begin{table}
\centering
\caption{Inference Latency and Throughput}
\begin{tabular}{c c c}
\hline
Model & Latency (ms) & Throughput (Hz) \\
\hline
Teacher model (8B) & 283.1 & 3.5 \\
Student model (1B, KD) & 129.2 & 7.7 \\
Base student model (1B) & 131.5 & 7.6 \\
\hline
\end{tabular}
\label{latency}
\end{table}
\subsection{KD-EKF Localization Performance Under LOS/NLOS Variations}
\label{fusion_comparison}

This section evaluates whether the proposed KD-EKF-based fusion localization algorithm maintains consistent performance across different indoor environment structures compared to conventional methods. To this end, comparative experiments were conducted in two structurally distinct indoor environments using the same UWB/PDR sensor configuration and EKF-based fusion framework. By varying the environmental layout and the distribution of LOS and NLOS regions, the environment generalization capability of the proposed algorithm is examined. Since the objective of this evaluation is to investigate whether EKF can adaptively respond to environmental changes through the incorporation of an LLM, the comparison is performed against a conventional EKF with fixed parameters. In all experiments, regions equipped with UWB anchors are defined as LOS areas, while regions without anchors are regarded as NLOS areas.

\subsubsection{Experiment 1: Short-Range Mixed LOS/NLOS Environment}
\begin{figure}
\centering
\includegraphics[width=0.85\linewidth]{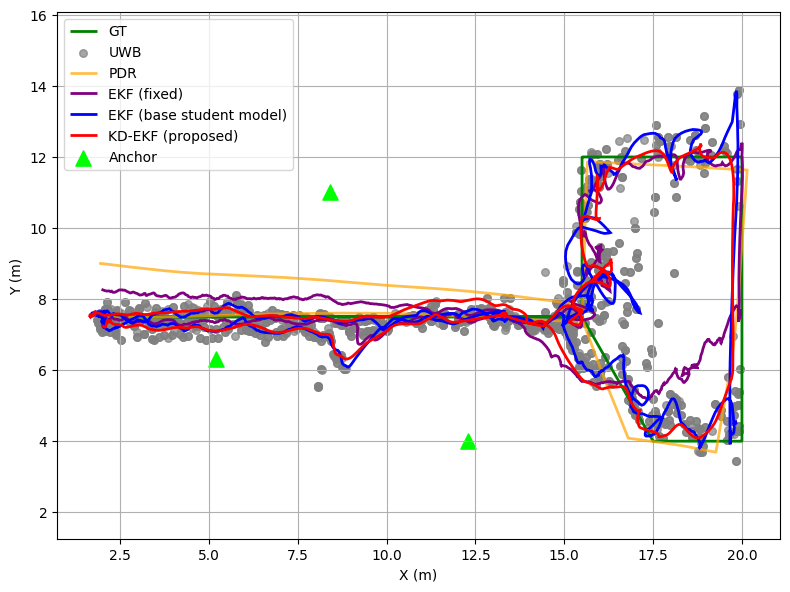}
\caption{Comparison of estimated trajectories for Experiment 1 in a short-range environment with mixed LOS/NLOS conditions.}
\label{exp1_traj}
\end{figure}

Fig.~\ref{exp1_traj} illustrates the trajectory comparison results for the GT, UWB-only localization, PDR-only estimation, EKF Fixed, the EKF (base student model), and the proposed KD-EKF. UWB-only localization exhibits large dispersion in NLOS segments, while PDR suffers from long-term drift due to cumulative errors. The fixed-parameter EKF reduces average localization error but fails to adapt effectively to NLOS-induced measurement degradation, leading to persistent trajectory deviation. The EKF (base student model) achieves a substantial improvement in fusion performance by significantly reducing both the mean error and the RMSE. The proposed KD-EKF further suppresses trajectory divergence and maintains close alignment with the GT trajectory, particularly in NLOS segments.

Table~\ref{exp1_ate} summarizes the quantitative ATE results for Experiment~1. The proposed KD-EKF achieves a mean error of 0.35 m, an RMSE of 0.45 m, and a maximum error of 0.85 m, outperforming the EKF Fixed with a mean of 0.70 m and an RMSE of 0.90 m, PDR with a mean of 0.85 m and an RMSE of 1.05 m, and UWB with a mean of 1.10 m and an RMSE of 1.35 m. The higher UWB error observed in this experiment is attributed to frequent NLOS-induced ranging biases caused by corridor reflections and structural occlusions, which significantly degrade trilateration accuracy. Meanwhile, short-term PDR motion estimation remains relatively stable over limited trajectories despite long-term drift accumulation. The EKF-based UWB/PDR fusion using the base student model significantly reduces localization error to a mean of 0.46 m and an RMSE of 0.60 m by adaptively regulating measurement covariance. Building upon this improvement, incorporating KD further refines reliability estimation and yields the lowest error achieved by the proposed KD-EKF. These results demonstrate that KD further enhances covariance adaptation beyond the learned student model, even in relatively simple mixed LOS/NLOS environments.

\subsubsection{Experiment 2: Closed Rectangular Corridor with Repeated LOS/NLOS Transitions}
Experiment~2 was conducted in a closed rectangular corridor environment with a centrally obstructed region, where the subject traversed the loop in a counterclockwise direction. The trajectory consists of four straight segments and four corner regions. Among the straight segments, two are equipped with UWB anchors forming LOS conditions, while the remaining segments and all corner regions lack anchor coverage and therefore exhibit persistent NLOS conditions. Consequently, the trajectory repeatedly alternates between LOS and NLOS regions.

\begin{figure}
\centering
\includegraphics[width=0.95\linewidth]{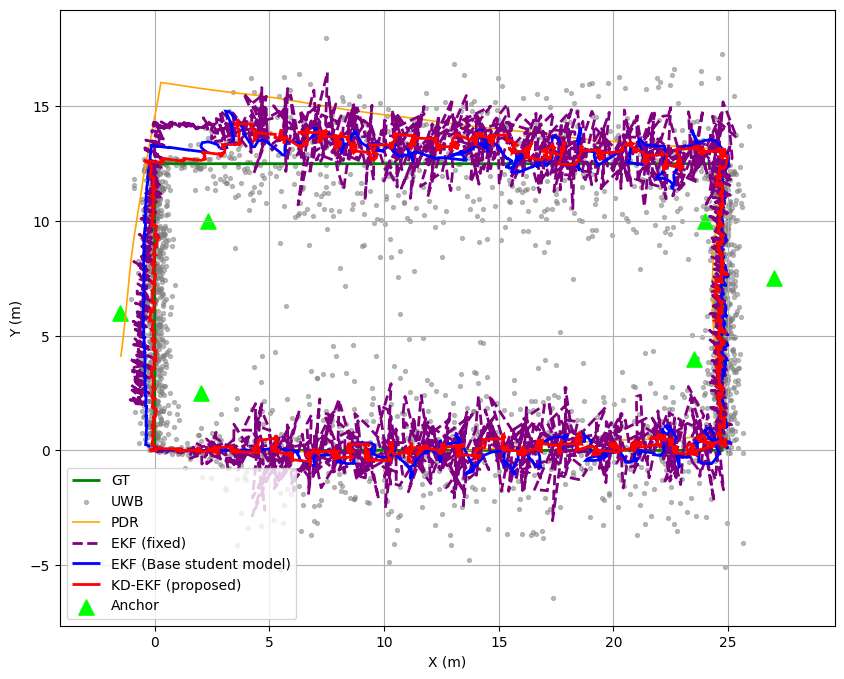}
\caption{Trajectory comparison for Experiment~2 in a closed rectangular corridor with repeated LOS/NLOS transitions.}
\label{exp2_traj}
\end{figure}

Fig.~\ref{exp2_traj} presents the trajectory comparison results for Experiment~2. The fixed-parameter EKF shows relatively stable performance in LOS segments but experiences abrupt error growth upon entering NLOS regions, with limited recovery afterward. PDR continues to accumulate drift even in the closed-loop trajectory, and UWB-only localization suffers from severe dispersion in NLOS segments. In contrast, the proposed KD-EKF effectively constrains error growth during repeated LOS/NLOS transitions and maintains consistent alignment with the GT trajectory throughout the loop.

Table~\ref{exp2_ate} presents the quantitative ATE results for Experiment~2, with overall performance trends generally consistent with those observed in Experiment~1. EKF-based UWB/PDR fusion with adaptive covariance regulation demonstrates clear accuracy improvements over fixed-covariance EKF-based UWB/PDR fusion and single-sensor approaches, confirming the effectiveness of dynamic reliability-driven sensor fusion.

In Experiment~1, PDR slightly outperformed UWB due to severe corridor-induced NLOS ranging biases that significantly degraded UWB trilateration accuracy. In Experiment~2, however, the error characteristics differ owing to the closed-loop trajectory structure. The repeated re-observation of LOS anchor configurations enables partial correction of accumulated UWB ranging biases, leading to improved average accuracy compared to PDR. Meanwhile, PDR drift continues to accumulate over the extended looped trajectory, resulting in higher mean errors despite relatively smooth short-term motion estimates. These findings indicate that single-sensor performance is strongly influenced by environmental geometry and trajectory configuration.

Building upon this fusion performance, the proposed KD-EKF further refines reliability estimation, leading to improved error stability and better control of peak errors. This indicates that KD does not replace the core structure of adaptive UWB/PDR fusion, but rather enhances it by providing an additional layer of performance refinement.
\begin{table}
\centering
\caption{Experiment~1: ATE performance comparison (Unit: \,m)}
\setlength{\tabcolsep}{15pt}
\begin{tabular}{l c c c}
\hline
Method & Mean & RMSE & Max \\
\hline
UWB & 1.10 & 1.35 & 2.50 \\
PDR & 0.85 & 1.05 & 1.40 \\
EKF (Fixed) & 0.70 & 0.90 & 1.80 \\
EKF (Base student model) & 0.46 & 0.6 & 1.45 \\
KD-EKF (Proposed) & \textbf{0.35} & \textbf{0.45} & \textbf{0.85} \\
\hline
\end{tabular}
\label{exp1_ate}
\end{table}
\begin{table}
\centering
\caption{Experiment~2: ATE performance comparison in a closed rectangular corridor environment (Unit: \,m)}
\setlength{\tabcolsep}{15pt}
\begin{tabular}{l c c c}
\hline
Method & Mean & RMSE & Max \\
\hline
UWB & 0.96 & 1.42 & 6.43 \\
PDR & 1.72 & 2.27 & 4.34 \\
EKF (Fixed) & 1.09 & 1.34 & 4.05 \\
EKF (Base student model) & 0.51 & 0.63 & 2.50 \\
KD-EKF (Proposed) & \textbf{0.36} & \textbf{0.49} & \textbf{1.97} \\
\hline
\end{tabular}
\label{exp2_ate}
\end{table}

Overall, the results from Experiments~1 and~2 demonstrate that the proposed KD-EKF is not a tuning-specific solution tailored to a single environment, but rather provides consistent performance improvements across indoor environments with different structures and LOS/NLOS distributions. This experimentally validates that integrating KD-based position prediction into EKF covariance adjustment effectively achieves environment generalization, which is a critical requirement for practical indoor localization systems.

\subsection{Ablation Study}
\label{ablation}

This section systematically analyzes the role of each component in the proposed KD-EKF framework from the perspective of dynamic measurement reliability and uncertainty estimation under time-varying indoor sensing conditions. The system is decomposed into functional elements responsible for reliability inference and covariance regulation, enabling a detailed evaluation of how each component contributes to probabilistic sensor fusion. Specifically, the covariance-based probabilistic integration of UWB and PDR, the adaptive measurement covariance scaling mechanism, and the KD-based reliability learning component are examined independently while preserving the overall fusion framework.

This analysis aims to verify that the performance improvements observed in Section~\ref{fusion_comparison} are not the result of environment-specific tuning or a single learning technique, but instead arise from explicitly modeling evolving measurement reliability and embedding this uncertainty into the EKF. In particular, the ablation analysis identifies real-time covariance adaptation as the primary mechanism through which reliability estimation translates into robust fusion performance.
 By isolating each component, the proposed framework demonstrates a principled approach for handling the non-stationary sensor quality variations commonly encountered in real indoor environments.

\subsubsection{Stability Improvement through Knowledge Distillation}

As shown in Table~\ref{abl}, the EKF (base student model) trained without KD exhibits larger trajectory-level error variation and higher peak errors, despite maintaining the same network architecture and inference pathway. The proposed KD-EKF reduces the RMSE from 0.63 m to 0.49 m while simultaneously suppressing large-error events reflected in the maximum error metric.

\begin{figure}
\centering
\includegraphics[width=0.9\linewidth]{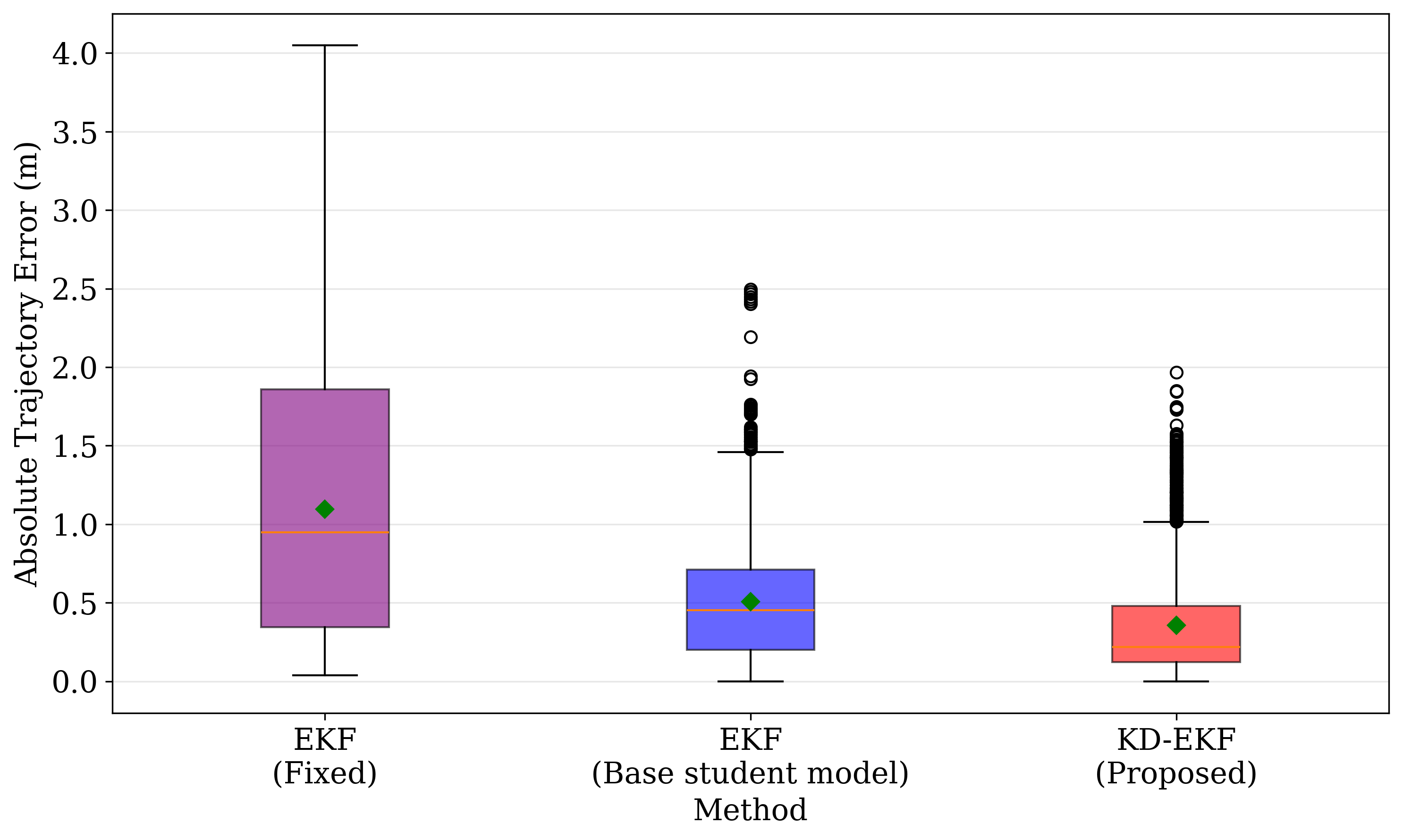}
\caption{Boxplot of ATE for the compared configurations.}
\label{abl_box}
\end{figure}

Figure~\ref{abl_box} further visualizes the ATE distribution for the ablation configurations in Experiment~1. Compared to the fixed EKF, the base student model significantly shifts the overall error distribution toward lower values due to adaptive covariance regulation. However, noticeable dispersion and long-tail outliers remain. With the incorporation of KD, the KD-EKF not only lowers the central tendency of the error distribution but also compresses the interquartile range and reduces extreme outliers. This distribution-level improvement indicates that KD enhances the stability of reliability estimation, preventing abrupt error amplification during LOS/NLOS transitions.

This indicates that KD enhances not merely point-wise regression accuracy but also the temporal consistency of reliability estimation. By transferring trajectory-level reasoning patterns from the teacher model, the student network becomes more responsive to gradual sensor degradation and abrupt NLOS transitions, thereby providing smoother and more informative reliability cues for the covariance regulation mechanism.
\subsubsection{Real-Time Integration of the Lightweight Reliability Estimator}

The large-scale teacher model is unsuitable for direct deployment in continuous EKF-based sensor fusion due to its high inference latency. In contrast, the student model trained via KD enables real-time reliability estimation while preserving the predictive structure learned by the teacher.

Notably, the base student model and the KD-trained student share identical computational complexity, confirming that the performance gains introduced by KD do not rely on increased model size or additional inference stages. This design allows learned uncertainty modeling to be incorporated into existing fusion pipelines without modifying system timing constraints or hardware requirements.
\begin{table}
\centering
\caption{Ablation study on the contribution of individual components in the proposed KD-EKF framework.}
\label{abl}

\small
\setlength{\tabcolsep}{4pt}
\renewcommand{\arraystretch}{0.9}

\begin{tabular}{lccc cc}
\hline
Configuration 
& KD 
& Student 
& Adaptive Cov. 
& RMSE (m) 
& Max (m) \\
\hline
EKF (Fixed)              
& -- & -- & -- & 1.34 & 4.05 \\

EKF (Base student model)        
& -- & \checkmark & \checkmark & 0.63 & 2.50 \\

KD-EKF (Proposed)       
& \checkmark & \checkmark & \checkmark & \textbf{0.49} & \textbf{1.97} \\
\hline
\end{tabular}

\normalsize
\end{table}

\subsubsection{Environmental Robustness via Adaptive Covariance Scaling}

The contribution of adaptive covariance regulation is isolated through comparison with a fixed-covariance EKF. As summarized in Table~\ref{abl}, the fixed EKF exhibits continuous error accumulation and elevated peak errors during sensor degradation and LOS/NLOS transitions.

In contrast, the proposed framework dynamically adjusts the measurement covariance based on learned reliability indicators, reducing the influence of low-confidence UWB observations while allowing greater reliance on PDR prediction when necessary. Even in conditions where the fixed EKF reaches an RMSE of 1.34 m, the KD-EKF maintains a stable performance level of 0.49 m, quantitatively demonstrating strong adaptability to diverse indoor structures and sensing conditions.

\section{Conclusion}
\label{con}
This paper reformulated indoor UWB/PDR fusion as a dynamic measurement reliability and uncertainty estimation problem, moving beyond EKF-based state estimation relying on fixed covariance parameters. By explicitly modeling NLOS-induced ranging degradation and inertial drift effects within the probabilistic covariance structure, the proposed KD-EKF framework enables fusion decisions to be governed by continuously evolving measurement trust rather than static noise assumptions.

Through KD, trajectory-level reliability reasoning learned by a large LLM-based teacher is transferred to a lightweight student model suitable for real-time deployment, allowing adaptive uncertainty modeling without environment-specific tuning. Central to this capability is the proposed real-time covariance adaptation mechanism, which continuously corrects EKF measurement uncertainty and fusion weighting using learned reliability indicators. This design enables the fusion system to respond dynamically to sensor degradation, multipath effects, and structural occlusions that arise in real indoor environments while eliminating the need for repeated manual recalibration.

Comprehensive experiments in structurally complex corridor environments demonstrate that the proposed framework consistently maintains stable localization performance across repeated LOS/NLOS transitions and long-duration motion sequences. Compared with fixed-covariance EKF-based UWB/PDR fusion and single-sensor approaches, the KD-EKF achieves lower average errors while effectively suppressing large-error events and long-term drift accumulation. These results confirm that explicitly estimating measurement reliability evolution is the central mechanism driving robust indoor localization, establishing dynamic uncertainty modeling as a principled foundation for scalable indoor positioning systems.

Future work will prioritize system-level optimization for scalable real-time deployment in industrial indoor environments. The integration of heterogeneous sensing modalities and contextual operational data will be explored to enable more comprehensive reliability-aware information fusion. In addition, tighter coupling between learned reliability inference and probabilistic filtering will be investigated to support long-term autonomous operation within complex industrial infrastructures.

\printcredits

\bibliographystyle{cas-model2-names}

\bibliography{cas-refs}



\end{document}